\documentclass[twocolumn]{aastex61}

\newcommand\aastex{AAS\TeX}

\submitjournal{ApJL}

\shorttitle{\aastex\ Molecular Gas in a $z=1.7$ BCG}
\shortauthors{Webb et al.}

\begin{document}

\title{Detection of a Substantial Molecular Gas Reservoir in a brightest cluster galaxy at $\MakeLowercase{z} = 1.7$}

\correspondingauthor{Tracy M.A. Webb}
\email{webb@physics.mcgill.ca}

\author{Tracy M.A. Webb}
\affil{McGill Space Institute 
Department of Physics 
McGill University  
3600 rue University 
Montr\'eal, Qu\'ebec, Canada, H3P 1T3}

\author{James Lowenthal }
\affiliation{Dept. of Astronomy, Smith College, Northampton, MA 01063, USA}


\author{Min Yun }
\affiliation{Department of Astronomy, University of Massachusetts, Amherst, Massachusetts, 01003 USA }

\author{Allison G.  Noble }
\affiliation{Kavli Institute for Astrophysics and Space Research, Massachusetts Institute of Technology, 77 Massachusetts Avenue, Cambridge, MA 02139, USA}

\author{Adam Muzzin}
\affiliation{Department of Physics and Astronomy, York University, 4700 Keele St., Toronto, Ontario, Canada, MJ3 1P3}

\author{Gillian Wilson}
\affiliation{Department of Physics and Astronomy, University of California-Riverside, 900 University Avenue, Riverside, CA 92521, USA }

\author{H.K.C. Yee}
\affiliation{Department of Astronomy \& Astrophysics, University of Toronto, 50 St. George St., Toronto, ON M5S 3H4, Canada}
\author{Ryan Cybulski}
\affiliation{Department of Astronomy, University of Massachusetts, Amherst, Massachusetts, 01003 USA }



\begin{abstract}

We report the detection of  CO(2-1) emission coincident with the brightest cluster galaxy (BCG) of the high-redshift galaxy cluster SpARCS1049+56, with the Redshift Search Receiver (RSR) on the Large Millimetre Telescope (LMT). We confirm a spectroscopic redshift for the gas of $z = $1.7091$\pm$0.0004, which is consistent with the systemic redshift of the cluster galaxies of $z = 1.709$.  The line is well-fit by a single component Gaussian with  a RSR resolution-corrected FWHM of 569$\pm$63 km/s.  We see no evidence for multiple velocity components in the gas, as might be expected from the multiple image components seen in near-infrared imaging with the Hubble Space Telescope. 
We measure the  integrated flux of the line to be 3.6$\pm$0.3 Jy km s$^{-1}$ and, using $\alpha_{\mathrm{CO}}$ = 0.8  M$_\odot$(K km s${-1}$ pc$^2$)$^{-1}$ we estimate a total molecular gas mass of 1.1$\pm$0.1$\times$10$^{11}$  M$_\odot$ and a M$_\mathrm{H2}$/M$_\star \sim$ 0.4;  This is the largest gas reservoir detected in a BCG above $z > 1$ to date.  Given the infrared-estimated star formation rate of 860$\pm$130 M$_\odot$yr$^{-1}$, this corresponds to a gas depletion timescale of $\sim$0.1Gyr.   We discuss several possible mechanisms for depositing such a large gas reservoir to the cluster center -- {\it e.g.,} a cooling flow, a major galaxy-galaxy merger or the stripping of gas from several galaxies --  but conclude that these LMT  data are not sufficient to differentiate between them.  
\end{abstract}



\section{Introduction}  \label{sec:intro}

Brightest cluster galaxies (BCGs) occupy special locations in the universe.  Because of this they likely experience evolutionary paths which are likewise special and unique, compared to other massive galaxies.  A picture has emerged whereby BCGs have grown in stellar mass by roughly a factor of 2 since $z\sim$1 \citep[e.g.,][]{lidman12,zhang16}, through the dry accretion of  satellite galaxies. In this scenario, most of the stellar mass of BCGs is produced at higher redshifts  in smaller systems, and assembled into the BCG at later times \citep{delucia07}.  Evidence is now mounting, however,  that beyond z$\sim$1, significant in-situ star formation may be occurring, rivalling the stellar mass build-up through dry mergers \citep{webb15b,mcdonald16}.   Although the occurrence of star formation in high-redshift BCGs is now established, the processes that fuel and trigger the star formation are poorly understood.  

At low redshifts, though rare, star forming BCGs primarily reside in cool-core clusters and the star formation rate is correlated with the cooling time of the X-ray gas \citep[e.g.,][]{rawle12}.  This appears to indicate that star formation in low-redshift BCGs is fuelled by large-scale cooling flows. Many of these systems also exhibit immense reservoirs of cold gas traced through the measurement of CO \citep[e.g.][]{edge01}. At higher redshift the situation is less clear, though  \citet{mcdonald14} has detected CO(3-2) in the Phoenix cluster with similar morphology \citep{russell16}.

Recently, however \citet{webb15b} and \citet{mcdonald16} have suggested that beyond $z\sim$ 1 star formation in BCGs is driven by a different process:  gas rich major mergers occurring at the center of galaxy clusters. This conclusion is motivated by the change in slope of the specific star formation rate with redshift \citep{mcdonald16} and one case study  of an apparent gas-rich BCG merger \citep{webb15a} in the galaxy cluster SpARCS1049+56. 



SpARCS1049+56 is a   $z = 1.7$ red-sequence-selected galaxy cluster with a star bursting core. This system is one of a larger sample of $ 0.3 < z < $ 1.8 clusters, drawn from the Spitzer Adaptation of the Red Sequence Cluster Survey (SpARCS) \citep{muzzin09,wilson09}, whose brightest cluster galaxies (BCGs) exhibit signs of intense star formation \citep[][Bonaventura et al., in press]{webb15b}.    It is spectroscopically confirmed through 27 cluster members \citep{webb15a} and the  richness-estimated mass  of M$_\mathrm{500kpc} \sim$ 4$\times$10$^{14}$M$_\odot$ is in good agreement with the weak lensing estimate of 3$\times$10$^{14}$ M$_\odot$ (J. Jee personal communication).  Its BCG is coincident with intense infrared emission of L$_\mathrm{IR}$ = 6.2$\times$10$^{12}$ L$_\odot$ with an AGN-corrected star formation rate (SFR) of 860 M$_\odot$ yr$^{-1}$.   The HST image has revealed optical morphology that is consistent with a major galaxy merger:  a long tidal tail with $>$ 10 UV-luminious clumps arranged like `beads-on-a-string'.   These clumps appear to originate within the stellar halo of the BCG and extend for 60 kpc. 

Here we present the detection of molecular gas in SpARCS1049+56 through a measurement of the CO (2-1) line with the Redshift Search Receiver (RSR) on the Large Millimeter Telescope (LMT).  To our knowledge, it is the first detection of molecular gas in a BCG beyond $z = 1$.   It offers a new opportunity to study the gas that fuels star formation in BCGs at high redshift.  In the following we present the basic properties of the gas reservoir and discuss various gas deposition mechanisms. 

Throughout we choose a flat $\Lambda$ cold dark matter cosmology with $H_\circ$ = 75 km s$^{-1}$ Mpc$^{-1}$, $\Omega_\mathrm{m} = 0.3$ and $\Omega_\Lambda = 0.7$. When required we adopt a Chabrier IMF.

\section{Observations} \label{sec:obs}

\subsection{Redshift Search Receiver Observations with the Large Millimeter Telescope}

  Observations were obtained using the Redshift Search Receiver (RSR) on the  32 meter Large Millimeter Telescope \citep{hughes10} on April 11-12, 2016 and February 18-19, 2017 for a total of 6 hours.  The wide bandwidth of the RSR covers the frequency window of 73--111 GHz simultaneously with a spectral resolution of 31.25 MHz (110 km~s$^{-1}$ at 85 GHz).  The redshifted CO (2--1) line ($\nu_0 = 230.5380$ GHz) at $z=1.7$ falls near the center of the spectrometer band at $\nu_{obs}\sim 85$ GHz.  Observing conditions were excellent with $\tau_{225GHz}=0.05-0.15$.  The final spectrum shown in Fig.\,\ref{fig:spec} was produced by calibrating and averaging the raw data using the facility data reduction software \textsc{DREAMPY}.

\section{Results}

\subsection{Detection and Characterization of the CO Line}

\begin{figure}
	\includegraphics[width=\columnwidth]{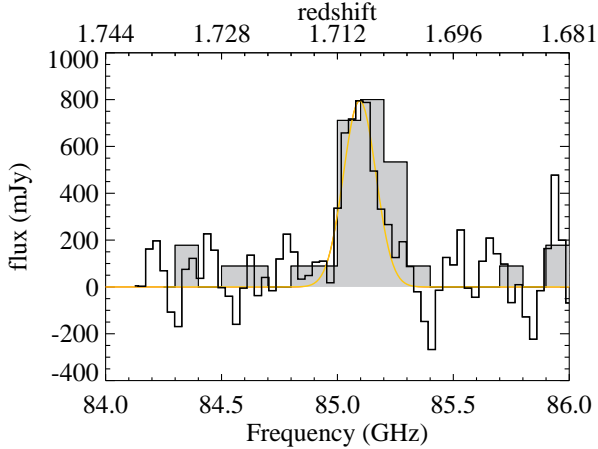}
    \caption{The Redshift Search Receiver spectrum showing the 10.7$\sigma$ detection of CO (2-1), redshifted to $z=1.7091$, solid black line. Over- plotted (solid yellow line) is the best fit Gaussian described in Section 3.1. Also shown are the locations in frequency space of the cluster galaxy members confirmed with optical spectroscopy (grey histogram).      }
    \label{fig:spec}
\end{figure}

Fig.~\ref{fig:spec} shows the RSR spectrum, zoomed in on CO(2-1) line. 
To characterize the line we  performed MCMC fitting of a Gaussian function on the data \citep{cybulski16}.  We fit over a 2 GHz bandwidth centered at 85 GHz,  with a low order baseline removal.    We measure the integral line flux to be 3.59$\pm$0.34 Jy km s$^{-1}$, providing detection significance of S/N = 10.7.  The line centroid is 85.0961$\pm$0.0131 GHz and it has a RSR-resolution corrected FWHM of 569$\pm$63 km s$^{-1}$.  

We calculate an integrated source brightness temperature of 1.16$\pm$0.10$\times$10$^{11}$ K km s$^{-1}$ pc$^2$ using:

\begin{equation}
L'_\mathrm{CO} = 3.25\times10^7 \times S_\mathrm{CO}\Delta v {D_L^2 \over {(1+z)^3 \nu_{obs}^2 }} \mathrm{K~km~s^{-1}~pc^2}
\label{eq:lum}
\end{equation}

\noindent where S$_\mathrm{CO}\Delta v$ is the integral line flux in Jy km s$^{-1}$, $D_L$ is the luminosity distance and $\nu_{obs}$ is the observed frequency.  

\begin{figure}

\includegraphics[width=\columnwidth]{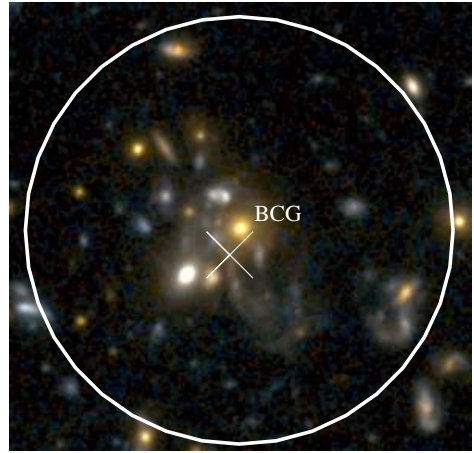}
\caption{The three-colour (F814W/F105W/F160W) HST image of the BCG and surrounding area. The large circle marks the size of the 25 arc second FWHM beam of the LMT. The LMT was pointed at the BCG (the central bright yellow galaxy in the image)  but the CO (2-1) emission may be coming from any galaxy within the beam, at $z = 1.709$. The location of the MIPS 24$\mu$m centroid is shown by the white cross, with the size of the cross corresponding to the 1$\sigma$ uncertainty in the MIPS position.  }
\label{fig:image}
\end{figure}

In Fig.~\ref{fig:image} we show  a zoom-in on the HST image of the BCG and its surroundings   \citep{webb15a}, with the LMT pointing position indicated. The LMT beam at 3mm (100 GHz) is $\sim$ 25 arcseconds  FWHM which encompasses the  $\sim$ 100 kpc surrounding the BCG, and all of the related substructure seen in Fig.~\ref{fig:image}.  Thus, the measured value for $L'_\mathrm{CO}$ should be taken as a lower-limit since, if the peak of the emission is actually off-axis, centered on one of the clumps in the tidal tail for example, the beam response function will result in an under-estimate of the emission.    
We note, however, that there is only one {\it Spitzer Space Telescope } MIPS source within the beam and it is within 2 arcseconds of the pointing center. 
This source is detected at a significance of 10$\sigma$. 
\subsection{Inferred Gas Properties}

The integrated source brightness temperature, determined in the previous section, can be used to estimate the total mass of molecular hydrogen \citep{solomon92}:
\begin{equation}
{M_\mathrm{H2} \over M_\odot}  = {\alpha_\mathrm{CO} \over r_{J1} } {L'_{(J-[J-1])} \over \mathrm{K~km~s}^{-1} ~ \mathrm{pc}^2}
\label{eq:mass}
\end{equation}

\noindent where $\alpha_\mathrm{CO}$ is the CO-to-H$_2$ conversion factor and   $r_{J1}$ provides a correction from the flux of the higher order J transitions to that of CO(1-0).   Here we  adopt an   $r_{21}$ value of 0.85 \citep{carilliwalter} and  $\alpha_\mathrm{CO}= 0.8 ~ \mathrm{M}_\odot$ (K km s$^{-1}$ pc$^2$)$^{-1}$; this is appropriate for ultraluminous infrared galaxies (ULIRGs)  and provides a conservative estimate on the total gas mass. We find a molecular gas mass of 1.1$\pm$0.1 $\times$10$^{11}$M$_\odot$. We note, however,  that, as reviewed by 
\citet{carilliwalter} and \citet{bolatto13},   the CO-to-H2 conversion is subject to some uncertainty, especially for star-forming systems and can be as high as   $\alpha_\mathrm{CO}\sim 4 ~ \mathrm{M}_\odot$ (K km s$^{-1}$ pc$^2$)$^{-1}$ for main sequence galaxies  \citep{decarli16} and \citet{daddi10aco}.

\section{Discussion} 

The precise line fitting provides us with a very accurate redshift for the CO gas of 1.7091$\pm$0.0004.  This is in excellent agreement with the systemic redshift of the galaxy cluster \citep{webb15a} of $z = 1.709\pm0.001$, estimated from an average of optical redshifts of 27 confirmed cluster members \citep{webb15a}.
This suggests that the CO emission is indeed coming from the center of the cluster potential, spatially near the BCG, and is not a projection along the line of sight from a galaxy on the outskirts of the cluster.

\begin{figure}
\includegraphics[width=\columnwidth]{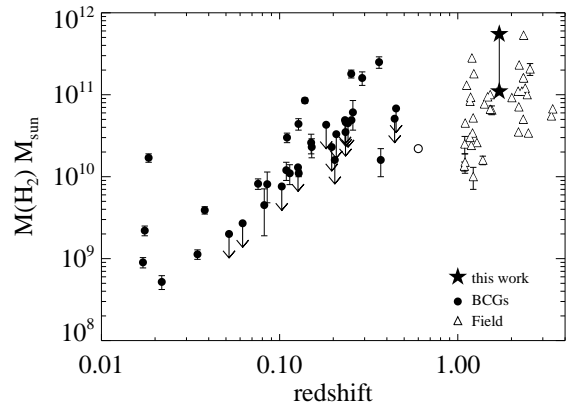}

\caption{A compilation of molecular gas masses, shown as a function of redshift. Note that as this is not a complete nor homogeneously selected sample, an evolutionary trend should not be drawn from these points. We show the values simply to place the molecular gas mass of this work in the context of other, similar measurements.  SpARCS1049 is denoted by the solid star. Solid circles correspond to BCGs from \citet{edge01,mcnamara14,russell14,vantyghem16}, with the open circle showing the location of the Phoenix BCG \citep{mcdonald13}.  Open triangles indicate star forming field galaxies from \citet{genzel10,decarli16}.  These studies have each chosen the  $\alpha_{CO}$ conversion factor deemed most appropriate for their respective systems, ranging from 0.8-4.0 M$_\odot$(K km s$^{-1}$ pc$^2$)$^{-1}$; in the text we use $\alpha_{CO}$ = 0.8 M$_\odot$(K km s$^{-1}$ pc$^2$)$^{-1}$, but on this plot we illustrate, with two points, the full range in mass that results from the range in $\alpha_{CO}$.    }
\label{fig:mass}
\end{figure}

Thus, these results indicate that a vast reservoir, 1.1$\pm$0.1 $\times$10$^{11}$M$_\odot$, of molecular gas has been deposited into the center of this galaxy cluster, and has formed the fuel source for a phase of rapid star formation.   To place this measurement  in context we show, in Fig.~\ref{fig:mass}, a compilation of molecular gas masses for comparable systems. We include the massive star-forming galaxies at $z > $ 1  from \citet{genzel10} and \citet{decarli16} as well as lower redshift ($z< 0.6$) BCGs \citep{edge01,mcdonald13,mcnamara14,russell14,vantyghem16}.  SpARCS1049 lies at the average of the mass distribution for CO detected systems.

In \citet{webb15a}, before the CO detection presented here, we speculated that the gas feeding the star formation might have been transported from the outskirts of the cluster  via a single gas-rich massive galaxy, due to the merger-like optical morphology of the system.  At low redshift such an event is  generally thought to be  difficult to achieve  due to the removal of the interstellar medium (ISM) of infallling galaxies by the intra-cluster medium (ICM). At high redshift, however, galaxies not only start their descent into the galaxy cluster potential with higher fractions of molecular gas, clusters themselves are younger and may have ICM's that are less effective at ISM removal than at low redshift.  While the major-merger picture is in line with other studies \citep{mcdonald16}, these data now raise several questions. 

The emission is well-fit by a single Gaussian and shows no sign of multiple velocity peaks.  This is perhaps unexpected given the multiple spatial components and merger-like morphology seen in the rest-frame optical (Fig.~2) and is also unlike many other CO-detected systems at high redshift. The sample of \citet{greve05}, for example, presents several cases (30-50\% of the sample) of dual-peaked CO spectra with velocity offsets of several hundreds km/s, which they conclude are produced through major mergers.   Similar dynamic structure is seen in the famous    local Antennae galaxies \citep{gao01,schulz07}.   The lack of such a signature in our data raises the possibility, but is not unambiguous evidence,  that the molecular gas, and thus the star formation, is not distributed over the same multiple objects as the optical/UV emission. Rather, it may be concentrated in a single dynamically distinct region.   

We posit two possible scenarios in which a gas-rich
merger is not required to produce the large amount of star formation
we see.  (1) The gas deposit may be the result of the stripping of many galaxies within the cluster center. In this case it may be that the multiple stellar clumps  have also come from several objects. Such a process would allow for a large build-up of gas which would have a velocity dispersion that traces the overall velocity dispersion of the cluster.  The velocity dispersion of the cluster is poorly constrained (430$^{+80}_{-100}$ km/s) but, given the existing velocities and its mass of $\sim$ 4$\times$10$^{14}$M$_\odot$ \citep{webb15a}, it is roughly a factor of  two larger than the velocity dispersion of the gas (Fig.\ref{fig:spec}, FWHM $\sim$560 km/s or $\sigma \sim$ 250 km/s.)  This is weak evidence against the idea that the gas has been stripped from many cluster members. (2)   The gas could flow to the center of the cluster through a large-scale cooling flow, as we see in many lower-redshift clusters (such as Phoenix \citep{mcdonald13}).  The complex optical morphology may indicate that cooling has been been triggered through the disruption of centralized heating processes via a collision between an infalling galaxy and the BCG (M.~Voit, personal communication).

\begin{figure}
		\includegraphics[width=\columnwidth]{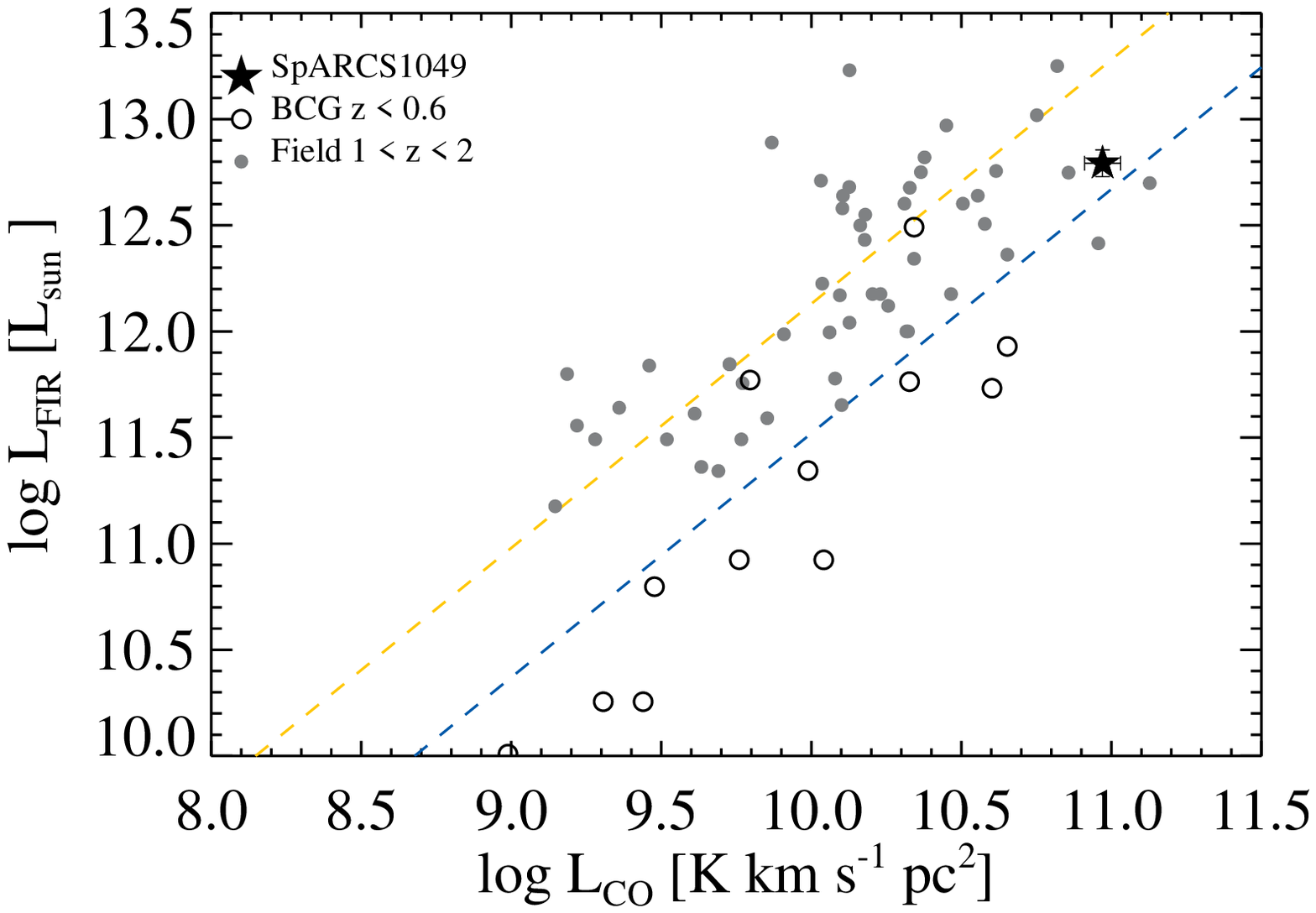}

    \caption{The relation between  L$_\mathrm{CO}$ and L$_\mathrm{FIR}$. Included are systems between 1 $<z<$ 2 with published CO measurements (grey points)
    \citep[][and references therein]{carilliwalter} and lower redshift ($z < 0.6 $) IR-bright BCGs \citep{edge01,odea08, rawle12}  We take the lowest CO transition available and correct it, following \citet{carilliwalter}, to CO(1-0).   The two dotted lines correspond to the best fit relations for main sequence galaxies (blue) and starburst galaxies (yellow) of \citet{daddi10blaws} and \citet{genzel10}.  The solid black star indicates the values measured for the SpARCS1049 BCG system.  }
    
    \label{fig:lcolir}
\end{figure}

Using our ancillary data we can attempt some interesting comparisons to measurements at other wavelengths in the hope they may shed light on the origin of the gas. Perhaps the simplest is the relation between the molecular gas luminosity (L$'_\mathrm{CO}$)  and the total infrared luminosity (L$_\mathrm{IR}$), since it is likely these emissions are  indeed spatially co-located.  Several sources of infrared imaging are available for this cluster:  {\it Herschel} Space Observatory, {\it Spitzer} Space Telescope and the James Clerk Maxwell Telescope.  We reiterate that in the highest resolution infrared imaging available ({\it Spitzer} Space Telescope MIPS)  there is only one infrared source within the 25 arc second  LMT beam, and therefore it is unlikely that the CO emission is coming from multiple sources separated by more than a few arc seconds.   The infrared luminosity was measured in \citet{webb15a} by fitting a template spectral energy distribution (SED) to the 24/250/350/450/500/850$\mu$m flux measurements.   In Fig.~\ref{fig:lcolir} we show the position of the SpARCS1049 BCG system on the L$'_\mathrm{CO}-$L$_\mathrm{IR}$ plane, compared to several other measurements of field galaxies at  $z > 1$ and lower redshift BCGS, drawn from the literature.     We see that although SpARCS1049 is extreme in its properties, in that it is relatively IR-  and CO-bright, it lies within the distribution of other infrared luminous galaxies at similar redshifts, inside and outside of clusters. 

These measurements indicate that even with such a large reservoir of molecular gas, the gas consumption timescale of SpARCS1049+56 remains rapid, $<\tau_{gas}>  $=  M$_\mathrm{H2}$/SFR $\sim$  0.1 Gyrs, because of the prodigious SFR. Although on the short end, this is within the normal range of local and high-redshift star forming galaxies \citep{saintonge11}. Concerning the origin of the gas, however, the  comparison has very little diagnostic power. The literature sample comprises a heterogeneous mixture of QSOs, HzRGs, SMGs, 24$\mu$m-selected, BzK, ERO,  and BMBX galaxies as well as cooling flow-fed BCGS.     Thus, despite the fact that these galaxies likely have different gas accretion mechanisms (mergers, cold-flows, cluster cooling flows) they exhibit similar  L$'_\mathrm{CO}-$L$_\mathrm{IR}$ properties.

\begin{figure}
\includegraphics[width=\columnwidth]{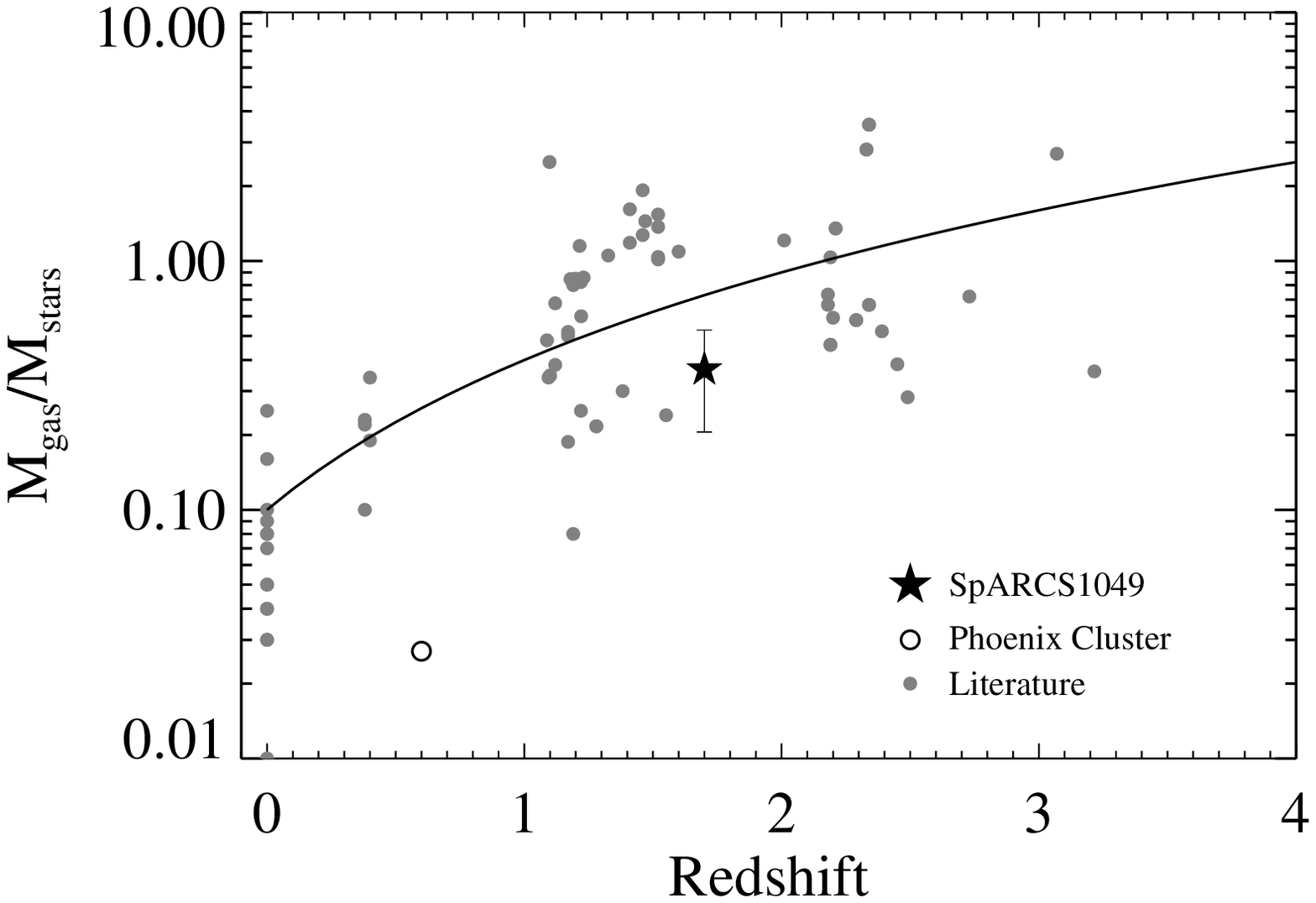}

\caption{  The ratio of molecular gas mass to stellar mass (M$_\mathrm{H2}/$M$\star$) for several star forming field galaxies from the literature (grey points) \citep{tacconi10,riechers10,magnelli12,daddi10aco,genzel10,leroy08,geach11,papovich16,silverman15,decarli16}.  The solid line corresponds to the relation from \citet{geach11}. As in Figure \ref{fig:mass} different $\alpha_{CO}$ values have been employed.  Nor have we attempted to normalize the stellar mass measurements to the same IMF (although the bulk of the studies adopt a Chabrier IMF). Because of this, scatter at the level of a factor of two is expected in the  M$_\mathrm{H2}/$M$\star$  measurements. }
\label{fig:ratio}
\end{figure}

The molecular-mass-to-stellar-mass  ratio (M$_\mathrm{H2}/$M$_\star$) is shown in Fig.\ref{fig:ratio}. This is a difficult quantity to estimate  because we do not know from which stellar mass component the CO originates. We adopt the stellar mass of the BCG \citep{webb15a} and see once again that this value  of M$_\mathrm{H2}/$M$_\star \sim$ 0.4 is in good agreement with that of other high-redshift ULIRGs.    We do not compare to the values of lower redshift BCGs in this case as the BCGs of \citet{edge01} do not have published stellar masses.  This agreement with field systems indicates that  if the source of the gas was a single accreted galaxy, it must have had a mass    within a factor of $\sim$ 5  the  BCG mass,  otherwise  M$_\mathrm{H2}/$M$_\star$ would be too high, in comparison to other similar systems.  Alternatively, this ratio could be explained by the accretion of several lower mass galaxies, each with a M$_\mathrm{H2}/$M$_\star \sim$ 0.4.

Understanding the role of the three possible gas deposition  processes -- major merger, multiple galaxies, cooling flow --  will clearly require additional observations, such as X-ray imaging or high-resolution integral field spectroscopy. 
Moreover the synthesis of these results with data at other wavelengths is not straightforward due to the confused morphology of the system and the large beam size of the facility used in this study, since  it is not possible to tell from these data physically where the CO emission originates (recall Fig.~2).   Because of this, higher resolution CO observations are imperative. Finally, although SpARCS1049+56 has been drawn from a larger sample of BCGs we do not yet know if it is representative, and thus more CO observations of high redshift BCGs are warranted. 

\section{Conclusions}

We report the 10.7$\sigma$ detection of molecular gas, via CO(2-1) observations with the LMT, in the center of the $z = 1.7$ galaxy cluster, SpARCS1049+56.   This is the first detection of molecular gas coincident with a BCG above $z > 1$.  We draw several conclusions: \\

\noindent $\bullet$ The high-precision redshift of the CO(2-1) line of $z=$ 1.7091$\pm$0.0004 indicates that the molecular gas is located at the center of the gravitational potential of the cluster. It is therefore unlikely that the CO emission is due to a cluster member seen along the line-of-sight to the BCG.\\

\noindent $\bullet$ The CO (2-1) line is well fit by a single Gaussian and shows no evidence for multiple velocity peaks.  Its broad width of FWHM 569$\pm$63 km/s ($\sigma = $ 250 km/s)  is nevertheless much narrower than the expected overall velocity dispersion of the cluster.  \\
 
\noindent $\bullet$ Although it lies on the upper-end of the relation, the  L$'_\mathrm{CO}-$L$_\mathrm{IR}$ ratio of the system is in good agreement with that of other galaxies at similar redshifts ($1 < z < 2$).  These galaxies include extreme systems such as the cooling-flow-fed star forming BCG of the Phoenix cluster and merging SMGs, to more `normal' BzK systems. \\

\noindent $\bullet$ We infer a molecular gas mass of M$_\mathrm{H2} =$ 1.1$\pm$0.4 $\times$10$^{11}$ M$_\odot$.    Employing the infrared-estimated SFR yields a gas depletion timescale of $\sim$ 0.1 Gyrs. Adopting the stellar mass of the BCG provides an  M$_\mathrm{H2}/$M$_\star$ value consistent with other high-redshift ULIRGs (M$_\mathrm{H2}/$M$_\star \sim 0.4$).  \\

\noindent $\bullet$ Although we posit several gas deposition mechanisms these data are not able to differentiate between them.  We conclude that more data are required to establish if the immense gas reservoir at the center of this system has been fed by a cooling flow, a major galaxy-galaxy merger or the stripping of gas from several galaxies.

\acknowledgments

This work would not have been possible without the long-term financial support from the Mexican Science and Technology Funding Agency, CONACYT (Consejo Nacional de Ciencia y Tecnolog\'{i}a) during the construction and early operational phase of the Large Millimeter Telescope Alfonso Serrano, as well as support from the the US National Science Foundation via the University Radio Observatory program, the Instituto Nacional de Astrof\'{i}sica, \'{O}ptica y Electr\'{o}nica (INAOE) and the University of Massachusetts, Amherst (UMass). TMAW ackowledges the support of an NSERC Discovery Grant. 
GW acknowledges financial support for this work from NSF grant AST-1517863 and from NASA through programs GO-13306, GO-13677, GO-13747, GO-13845/14327 from the Space Telescope Institute, which is operated by AURA, Inc., under NASA contract NAS 5-26555. 
%

\vspace{5mm}
\facilities{LMT, HST, Spitzer, Herschel, JCMT, Keck}








\end{document}